\documentclass[aps,preprint,floats,epsf,epsfig,nofootinbib,letter]{revtex4}
\usepackage{epsfig,graphicx,color,appendix}

\usepackage{dcolumn}
\usepackage{bm}

\def\be{\begin{eqnarray}}
\def\en{\end{eqnarray}}
\def\non{\nonumber}

\def\CP{{\it CP}~}

\def\dg{{e^{-i\gamma}}}
\def\dpg{{e^{+i\gamma}}}
\def\da{{e^{i\delta_{A}}}}
\def\dc{{e^{i\delta_{C}}}}
\def\de{{e^{i\delta_{E}}}}
\def\dpp{{e^{i\delta_{P}}}}
\def\dpa{{e^{i\delta_{PA}}}}
\def\dpe{{e^{i\delta_{PE}}}}

\begin{document}


\vskip 5.5 cm

\title{ Model-Independent Analysis of CP Violation in Charmed Meson Decays }

\author{ Rohit Dhir$^a$\footnote{dhir.rohit@gmail.com}, C. S. Kim$^a$\footnote{Corresponding author,~ cskim@yonsei.ac.kr} and
Sechul Oh$^b$\footnote{scohph@yonsei.ac.kr}}

\address{$^a$Department of Physics \& IPAP,  Yonsei University,  Seoul 120-749, Korea \\
         $^b$University College, Yonsei University, Incheon 406-840, Korea}

\vspace*{0.5cm}


\begin{abstract}
\noindent We present a model-independent analysis of CP violation, inspired by recent experimental observations, in charmed meson decays. The topological diagram approach is used to study direct CP asymmetries for singly Cabibbo-suppressed two-body hadronic decays of charmed mesons. We extract the magnitudes and relative phases of the corresponding topological amplitudes from available experimental information. In order to get more precise and reliable estimates of direct CP asymmetries, we take into account  contributions from all possible strong penguin amplitudes, including the internal $b$-quark penguin contributions. We also study flavor SU(3) symmetry breaking effects in these decay modes and consequently, predict direct CP asymmetries of unmeasured modes.
\\ \\
$*$ Keywords: CP violation, Asymmetry, Charm, D meson, Topological diagram, SU(3) symmetry 
\\ \\
$*$ PACS numbers: 11.30.Er, 11.30.Hv, 13.25.Ft, 14.40.Lb  
\end{abstract}

\maketitle

%
%

\section{Introduction}

Numerous studies of CP violation have been carried out in quark flavor physics. The $b$- and $s$-quark sectors have provided a fertile testing ground for the the standard model (SM) explanation of CP violation through the Cabibbo-Kobayashi-Maskawa (CKM) matrix. B factory experiments played a central role to confirm the CKM framework and to determine each matrix element.

In the charm sector, there are stark differences for the study of CP violation.  Mixing occurs at extremely small rate in neutral charmed mesons, compared to that in neutral K and B mesons.  It results in the mixing-induced indirect CP asymmetry being  negligible.
For direct CP violation, the asymmetry can vary greatly depending on particular charmed meson decay modes. The largest direct CP asymmetries are expected in singly Cabibbo-suppressed (SCS) decays, such as $D^0 \to \pi^+ \pi^-$ and $D^0 \to K^+ K^-$, where interference between penguin and tree contributions can be substantial.
Naturally both theoretical and experimental interest has been focused on this type of charm decays.
In Cabibbo-favored (CF) decays, such as $D^0 \to K^- \pi^+$ and $D^+ \to \bar K^0 \pi^+$, the favored tree contribution dominates so that direct CP violation is negligible. In doubly Cabibbo-suppressed (DCS) decays, such as $D^0 \to K^+ \pi^-$ and $D^+ \to K^0 \pi^+$, direct CP asymmetries are expected to be negligible in the SM, but non-negligible in certain new physics (NP) models.

Three years ago, LHCb had observed indications of direct CP asymmetry at $3.5\sigma$ in $\Delta a_{CP}= a_{CP}(D^0 \to K^+ K^-) -a_{CP}(D^0 \to \pi^+ \pi^-)$~\cite{Aaij:2011in}. CDF and Belle also reported similar results~\cite{Collaboration:2012qw,Ko:2012px}, giving a world average of $\Delta a_{CP} = (-0.678 \pm 0.147)\%$.
These results attracted great attention and led to numerous theoretical works~\cite{Feldmann:2012js,Brod:2012ud,Brod:2011re,Cheng:2012wr,Cheng:2012xb,Li:2012cfa,Bhattacharya:2012ah,Franco:2012ck,Isidori:2011qw,Grossman:2012ry,Atwood:2012ac,Giudice:2012qq,Hiller:2012wf,
Hochberg:2011ru,Altmannshofer:2012ur,Isidori:2012yx,Hiller:2012xm,Buccella:2013tya,Muller:2015lua,Biswas:2015aaa}, both in the context of the SM and of various models of NP.
For a long time, direct CP violation in charm decays was expected to be below $\mathcal{O} (10^{-3})$. However, some of the recent studies indicate that asymmetries at $\mathcal{O} (10^{-3})$ in these final states may not be excluded within the SM.
Since then, LHCb has reported new results from which the world average has moved much closer to zero: $\Delta a_{CP} = (-0.253 \pm 0.104)\%$~\cite{Amhis:2012bh}.
The direct CP asymmetries, reported by LHCb~\cite{Aaij:2014gsa}, for $D^0 \to \pi^+ \pi^-$ and $D^0 \to K^+ K^-$ are $a_{\rm CP} (\pi^+\pi^-) = (-0.20 \pm 0.19 ({\rm stat}) \pm 0.10 ({\rm syst}))\%$ and $a_{\rm CP} (K^+K^-) = (-0.06 \pm 0.15 ({\rm stat}) \pm 0.10 ({\rm syst}))\%$.
More precise measurements from future experiments, such as upgraded LHCb, BESIII and Belle II, are crucial to provide necessary information.
Besides CP asymmetries, the very different experimental values of branching fractions ${\cal B} (D^0 \to K^+ K^-)$ and ${\cal B}(D^0 \to \pi^+ \pi^-)$ have been a long-standing puzzle. Some efforts have been made to resolve the issue within the SM~\cite{Li:2012cfa,Biswas:2015aaa,Cheng:2010ry}.

For hadronic charmed meson decays, there is still no proper theoretical description of the underlying mechanism based on QCD. As well known, it is mostly because of the charm quark mass of order 1.5 GeV, which is not heavy enough to apply for a heavy quark expansion, and not sufficiently light to allow for the application of a chiral expansion.
Thus, for hadronic $D$ decays, it is not reliable to use the QCD-inspired theoretical approaches satisfactorily worked for $B$ meson decays, such as the QCD factorization (QCDF) approach~\cite{Beneke:2000ry,Beneke:1999br}, the perturbative QCD (pQCD) approach~\cite{Keum:2000ph,Keum:2000wi} and the soft-collinear effective theory~\cite{Bauer:2001yt}.

Since there is no reliable theoretical framework for hadronic decays of charmed mesons, a model-independent method, called the quark diagram approach~\cite{Chau,CC86,CC87}, has been developed. In this approach all the decay amplitudes are decomposed into the so-called topological amplitudes corresponding to the different topological quark diagrams.
Based on flavor SU(3) symmetry, the heavy meson decay amplitudes also can be decomposed into linear combinations of the SU(3) amplitudes which are SU(3) reduced matrix elements~\cite{Zeppenfeld:1980ex,SU3_Savage,Gronau:1994rj,Gronau:1995hm,Gronau:1995hn,SU3_Oh2,Oh:1998wa,SU3_He}.
This approach is equivalent to the quark diagram approach when flavor SU(3) symmetry is imposed to the latter.
Since each topological quark diagram includes all possible strong interactions to all orders, analyses of topological diagrams can provide information on final-state interactions (FSIs).
In this model-independent analysis, one can determine each topological amplitude from experimental data with the help of fitting, if a sufficiently large number of measurements are available.
So far several works have been done to study hadronic charmed meson decays in the framework of the quark diagram approach.
However, because of the difficulty to manage the large number of parameters in fitting, those works have used at least in part  certain model-dependent information ({\it e.g.,} information on SU(3) breaking from factorization or QCDF, etc.) to obtain a fit.

In this work we shall study direct CP asymmetries of SCS $D_{(s)} \to PP$ ($P = \pi, K, \eta^{(\prime)}$) decays using the updated experimental data.
For a least model-dependent analysis of the charmed meson decays, we choose the quark diagram approach and perform the $\chi^2$ analysis in the most general way, {\it i.e., without using any model-dependent information}~\footnote{This will be Case I of our analysis. In Case II we shall impose certain constraints on two of the parameters encouraged by the fit obtained for CF charmed meson decays. }.
The present experimental data show that the measured values of branching fractions of $D_{(s)} \to PP$ are quite accurate, but direct CP asymmetries include very large errors. In this situation our strategy is: (i) first, to fit the experimental data, especially the accurately measured branching fractions, by using the topological amplitudes as parameters, (ii) then, to extract each topological amplitude from the fit, and (iii) to make predictions for the direct CP asymmetries.
Since the large number of parameters are involved in this analysis, obtaining a satisfactory fit without using any model-dependent information turns out to be a very difficult task.

We would like to emphasize that (as we shall see later) for direct CP asymmetries of SCS D decays, the penguin amplitudes corresponding to the internal $b$-quark ({\it i.e.,} internal $b$-quark penguin amplitudes) multiplied by the CKM factor $V_{cb}^* V_{ub}$ become important so that it cannot be ignored.
In our analysis we shall explicitly express all the relevant $b$-quark penguin amplitudes, including the $b$-penguin exchange and the $b$-penguin annihilation ones. Then, we shall determine the magnitudes and strong phases of these $b$-quark penguin amplitudes.
This is one of different points of our work from the previous other works. In other words, all possible strong penguin contributions including the internal $b$-quark penguins are explicitly included in our analysis.
We shall see that the internal $b$-quark penguin contributions are comparable with the internal $s$- and $d$-quark penguin ones.

This paper is organized as follows. In Sec.~II, we introduce the Wolfenstein parametrization of the CKM matrix up to order $\lambda^6$ and topological quark diagrams relevant to $D \to PP$ decays. In Sec.~III, the explicit SU(3) decomposition of the decay amplitudes and its relevance to direct CP asymmetries are presented.
In Sec.~IV, we perform the $\chi^2$ analysis by taking into account SU(3) breaking to determine the topological amplitudes, and predict the direct CP asymmetries. Our conclusions are given in Sec.~V.

\section{Framework}

It is well known that the Wolfenstein parametrization of the CKM matrix can be easily obtained from the standard
Chau-Keung(CK) parametrization.
The CKM matrix elements up to order $\lambda^6$ are given by~\cite{Ahn:2011fg}
\begin{widetext}
\be   \label{eq:Wolf_CK_HO}
V_{\rm Wolf}^{{\rm (CK)}} &=&
\footnotesize
 \left( \begin{array}{ccc}
  1 -\frac{\lambda^2}{2} -\frac{\lambda^4}{8}
    & \lambda  ~,
    & A \lambda^3 (\rho -i \eta)
  \\
    -\frac{\lambda^6}{16} [1 +8 A^2 (\rho^2 +\eta^2)] ~,
    &
    &
  \\ \\
  -\lambda +\frac{\lambda^5}{2} A^2 (1 -2\rho -2i \eta) ~,
    & 1 -\frac{\lambda^2}{2} -\frac{\lambda^4}{8} (1 +4 A^2)
    & A \lambda^2
  \\
    & ~ -\frac{\lambda^6}{16} [1 -4 A^2 (1 -4\rho -4i\eta)] ~,
    &
  \\ \\
  A \lambda^3 (1 -\rho -i \eta)
    & -A \lambda^2 +\frac{\lambda^4}{2} A (1 -2\rho -2i \eta)
    & 1 -\frac{\lambda^4}{2} A^2
  \\
  +\frac{\lambda^5}{2} A (\rho +i \eta) ~,
    & +\frac{\lambda^6}{8} A ~,
    & -\frac{\lambda^6}{2} A^2 (\rho^2 +\eta^2)
 \end{array} \right)  +{\cal O}(\lambda^7) ~. \nonumber \\
\en
\end{widetext}
Regarding charm decays, the relevant CKM factors are $\lambda_q \equiv V_{cq}^* V_{uq}~ (q = d, s, b)$.
From the above matrix elements, one finds
\be   \label{eq:CKMelement1}
\lambda_d &=& -\lambda +\frac{\lambda^3}{2} +\frac{\lambda^5}{8} (1 +4 A^2) -\lambda^5 A^2 (\rho +i \eta) +{\cal O}(\lambda^7)
  \equiv \lambda_d^{(1)} +\lambda_d^{(2)} +{\cal O}(\lambda^7),
  \nonumber \\
\lambda_s &=& \lambda -\frac{\lambda^3}{2}  -\frac{\lambda^5}{8} (1 +4 A^2) +{\cal O}(\lambda^7),  \nonumber \\
\lambda_b &=& \lambda^5 A^2 (\rho -i \eta) +{\cal O}(\lambda^7) \equiv |\lambda_b| e^{-i\gamma} +{\cal O}(\lambda^7),
\en
where $\lambda_d^{(1)} = -\lambda +\frac{\lambda^3}{2} +\lambda^5 \left[\frac{1}{8} +\frac{1}{2} A^2 \right]$ and
$\lambda_d^{(2)} = -\lambda^5 A^2 (\rho +i \eta) \equiv -|\lambda_d^{(2)}| e^{+i\gamma} = -|\lambda_b| e^{+i\gamma}$.
Notice that the imaginary terms appear in both $\lambda_d$ and $\lambda_b$ at order $\lambda^5$~\footnote{ Note that up to order of $\lambda^6$, $\lambda_d \sim -\lambda -\lambda^5 e^{+i \gamma}$, $\lambda_s \sim \lambda$, and $\lambda_b \sim \lambda^5 e^{-i \gamma}$. Therefore, ${{|\lambda_b|} \over {|\lambda_{s (d)}|}} \sim \lambda^4 \sim 2 \times 10^{-3}$.
 }.
Thus, for CP asymmetries in charm decays, the internal $b$-quark contributions (penguin contributions) with the CKM factor $\lambda_b$ as well as the $d$-quark ones with $\lambda_d$ become important.

In SU(3) decomposition of the decay amplitudes for $D \to PP$ ($P = \pi, K, \eta^{(\prime)}$) modes, the decay amplitudes are expressed in terms of topological quark diagram contributions.
The topological amplitudes corresponding to different topological quark diagrams, as shown in
Figure 1, can be classified into three distinct groups as follows:
\begin{enumerate}
\item tree and penguin amplitudes:
\begin{itemize}
\item $\cal T$, color-allowed tree amplitude (equivalently, external $W$-emission);
\item $\cal C$, color-suppressed tree amplitude (equivalently, internal $W$-emission);
\item $\cal P$, QCD-penguin amplitude;
\item $\cal S$, singlet QCD-penguin amplitude involving SU(3)-singlet mesons (e.g., $\eta^{(\prime)}, ~\omega, ~\phi$);
\item $\cal P_{\rm EW}$ : color-favored EW-penguin amplitude;
\item $\cal P_{\rm EW}^C$ : color-suppressed EW-penguin amplitude;
\end{itemize}
\item weak annihilation amplitudes:
\begin{itemize}
\item $\cal E$, $W$-exchange amplitude;
\item $\cal A$, $W$-annihilation amplitude;
($E$ and $A$ are often jointly called ``weak annihilation'' amplitudes.)
\item $\cal PE$, QCD-penguin exchange amplitude;
\item $\cal PA$, QCD-penguin annihilation amplitude;
\item $\cal PE_{\rm EW}$ : EW-penguin exchange amplitude;
\item $\cal PA_{\rm EW}$ : EW-penguin annihilation amplitude;
($\tilde P \tilde E$ and $\tilde P \tilde A$ are also jointly called ``weak penguin annihilation''.)
\end{itemize}
\item flavor-singlet weak annihilation amplitudes: all involving SU(3)$_{\rm F}$-singlet mesons,
\begin{itemize}
\item $\cal SE$, singlet $W$-exchange amplitude;
\item $\cal SA$, singlet $W$-annihilation amplitude;
\item $\cal SPE$, singlet QCD-penguin exchange amplitude;
\item $\cal SPA$, singlet QCD-penguin annihilation amplitude;
\item $\cal SPE_{\rm EW}$ : singlet EW-penguin exchange amplitude;
\item $\cal SPA_{\rm EW}$ : singlet EW-penguin annihilation amplitude.
\end{itemize}
\end{enumerate}
The reader is referred to Ref.~\cite{Cheng:2011qh} for details.
Each topological quark diagram in this approach includes all possible strong interactions to all orders. Therefore, analyses of topological diagrams can provide information on final-state interactions.

\begin{figure}[t]
\vspace*{1ex}
\includegraphics[width=2.8in]{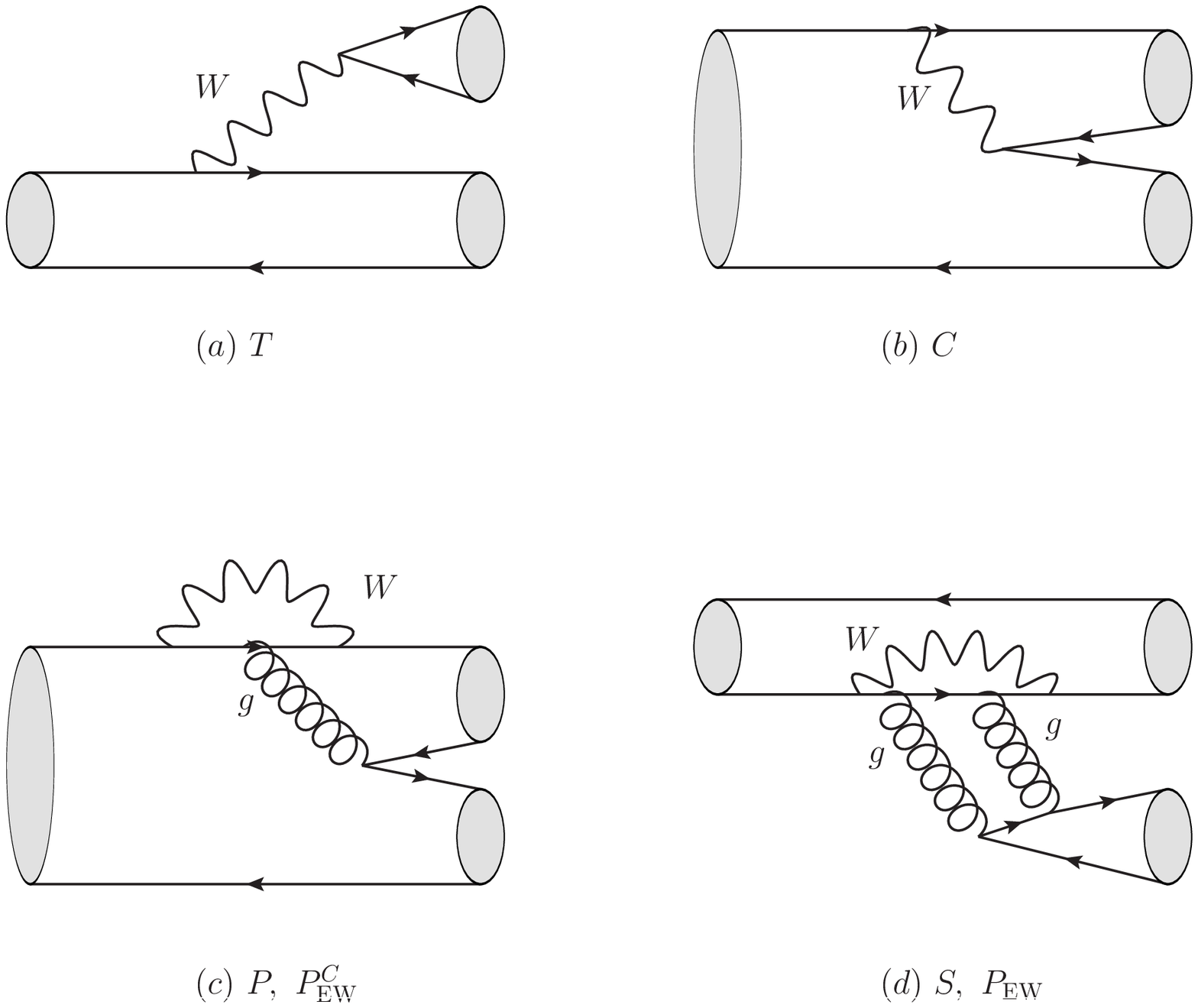}\qquad\includegraphics[width=2.8in]{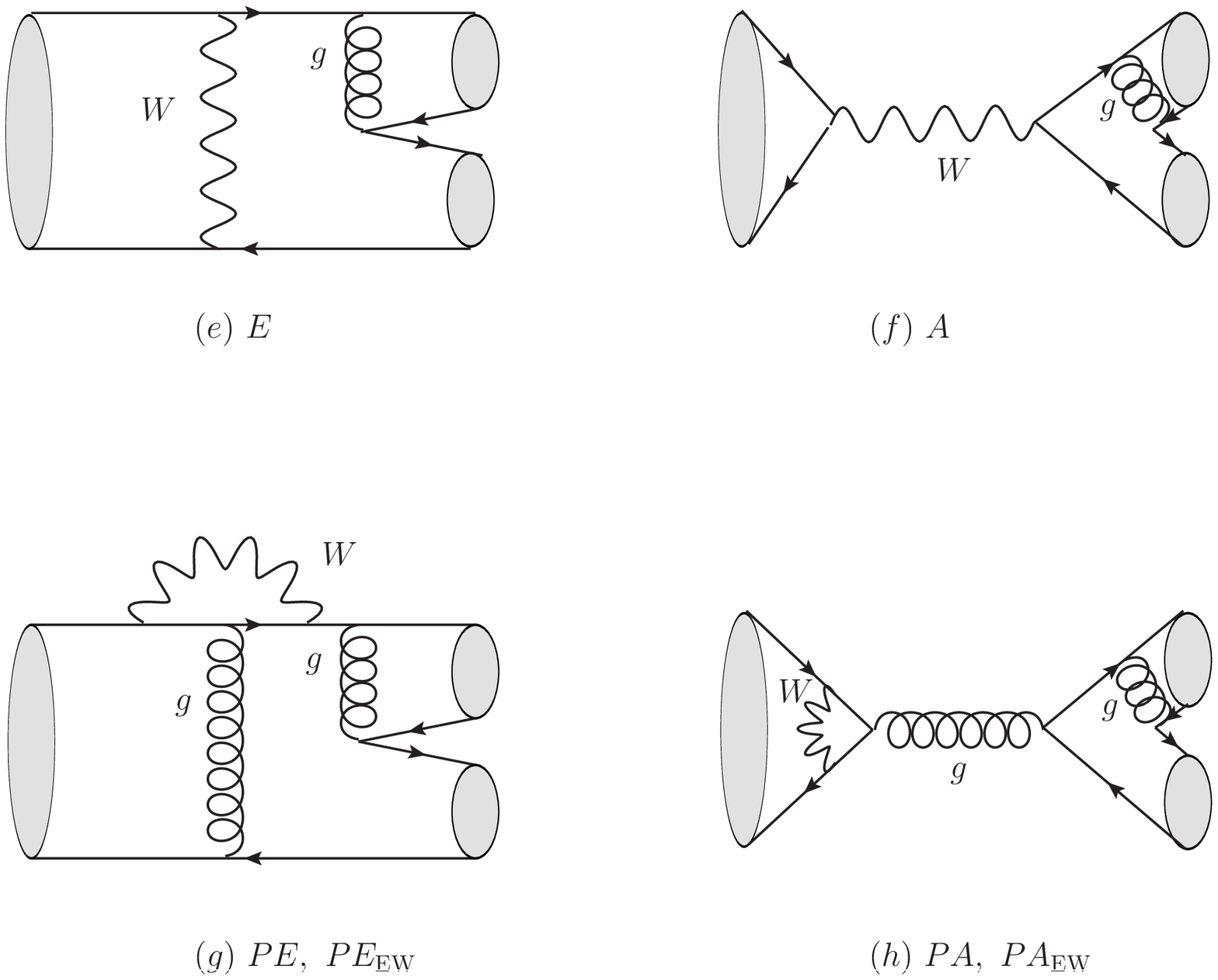}
\vspace*{-1ex}
\caption{ Topology of possible flavor diagrams ($not$ Feynman diagrams: Each topological diagram includes all possible strong interactions to all orders.):
(a) color-allowed tree $\cal T$, (b) color-suppressed tree $\cal C$,
(c) QCD-penguin $\cal P$, (d) singlet QCD-penguin $\cal S$ diagrams, (e) $W$-exchange $\cal E$, (f) $W$-annihilation $\cal A$, (g) QCD-penguin exchange $\cal PE$, and
(h) QCD-penguin annihilation $\cal PA$ diagrams.
The color-suppressed EW-penguin $\cal P_{\rm EW}^C$, color-favored EW-penguin $\cal P_{\rm EW}$, EW-penguin exchange $\cal PE_{\rm EW}$ and EW-penguin annihilation $\cal PA_{\rm EW}$ diagrams are obtained from proper replacement of gluon lines by $Z$-boson or photon lines in (c), (d), (g), (h), respectively.
 } \label{Fig:Quarkdiagrams}
\label{Fig1}
\end{figure}

\section{Decay amplitudes and CP asymmetries}

The decay amplitudes of two-body hadronic $D$ decays can be represented in terms of the topological quark diagram contributions.
In general the decay amplitudes of $D^0 \to \pi^+ \pi^-$ and $D^0 \to K^+ K^-$ can be written as
\be   \label{eq:Amp}
{\cal A} (D^0 \to \pi^+ \pi^-) &=& \lambda_d ({\cal T} + {\cal E} + {\cal P}_d + {\cal PA}_d + {\cal PE}_d)^{\pi\pi} + \lambda_s ({\cal P}_s + {\cal PA}_s + {\cal PE}_s)^{\pi\pi} \nonumber \\
      &+& \lambda_b ({\cal P}_b + {\cal PA}_b + {\cal PE}_b)^{\pi\pi};  \\
{\cal A} (D^0 \to K^+ K^-) &=& \lambda_s ({\cal T} + {\cal E} + {\cal P}_s + {\cal PA}_s + {\cal PE}_s)^{KK} + \lambda_d ({\cal P}_d + {\cal PA}_d + {\cal PE}_d)^{KK} \nonumber \\
      &+& \lambda_b ({\cal P}_b + {\cal PA}_b + {\cal PE}_b)^{KK} ,
\en
where ${\cal P}_q$, ${\cal PA}_q$, ${\cal PE}_q$ are the QCD-penguin, QCD-penguin annihilation, QCD-penguin exchange amplitudes with an internal $q$-quark, respectively.  Here the internal $b$-quark contributions with the CKM factor $\lambda_b$ are explicitly shown.~\footnote{For simplicity, electroweak-penguin contributions have been neglected since they are expected to be very small, as previous works in the literature.}

The possible sources of CP violation in charm transitions include: CP violation in $\Delta C=1$ decay amplitudes (direct CP violation) and CP violation through $\Delta C=2$ $D^0-\bar D^0$ mixing (Indirect CP violation). The latter has been estimated to be zero in recent experiments~\cite{Amhis:2012bh}.
The direct CP asymmetry is defined as
\be   \label{eq:DirectCP1}
a_{CP} (D \to f) = {{ {\cal B} (D \to f) - {\cal B} (\bar D \to \bar f) } \over { {\cal B} (D \to f) + {\cal B} (\bar D \to \bar f) }},
\en
where ${\cal B}$ denotes a branching fraction.  For $D \to f$ decay having two strong and two weak phases, the decay amplitude can be generally written as
\be   \label{eq:DecayAmp}
{\cal A} (D \to f) = |{\cal A}_1| e^{-i \delta_1} e^{-i \phi_1}+|{\cal A}_2| e^{-i \delta_2} e^{-i \phi_2} ,
\en
where $\delta_{1,2}$ and $\phi_{1,2}$ are strong and weak phases, respectively.  Then, the direct CP asymmetry is given by
\be   \label{eq:DirectCP2}
a_{CP} (D \to f) = {{ 2 |{\cal A}_1| |{\cal A}_2| \sin (\delta_1 - \delta_2) \sin (\phi_1 - \phi_2) } \over {|{\cal A}_1|^2 +|{\cal A}_2|^2 +2 |{\cal A}_1| |{\cal A}_2| \cos (\delta_1 - \delta_2) \cos (\phi_1 - \phi_2) }},
\en
which is non-zero only if $\Delta\delta = \delta_1 - \delta_2  \neq 0,~ \Delta\phi = \phi_1 - \phi_2 \neq 0$, and $|A_{1, 2}|  \neq 0$.

To study the direct CP asymmetry, the decay amplitude of $D^0 \to K^+ K^-$, for instance, can rewritten as
\be   \label{eq:DtoKK1}
{\cal A} (D^0 \to \pi^+ \pi^-) &=& {\cal A}_1^{\pi\pi} + {\cal A}_2^{\pi\pi},
\en
where up to order $\lambda^6$,
\be   \label{eq:Dtopipi1}
{\cal A}_1^{\pi\pi} &=& \lambda_d^{(1)} (T + E \de + P_d \dpp + PA_d \dpa + PE_d \dpe)  \nonumber \\
  && + \lambda_s (P_s \dpp + PA_s \dpa + PE_s \dpe)  \nonumber \\
  &\equiv& |{\cal A}_1^{\pi\pi}| e^{-i \delta_1},  \nonumber \\
{\cal A}_2^{\pi\pi} &=& \lambda_d^{(2)} (T + E \de + P_d \dpp + PA_d \dpa + PE_d \dpe)   \nonumber \\
  && + \lambda_b (P_b \dpp + PA_b \dpa + PE_b \dpe)   \nonumber \\
  &=& - |\lambda_d^{(2)}| (T + E \de + P_d \dpp + PA_d \dpa + PE_d \dpe) \dpg  \nonumber \\
  && + |\lambda_b| (P_b \dpp + PA_b \dpa + PE_b \dpe) \dg   \nonumber \\
  &\equiv& |{\cal A}_{2a}^{\pi\pi}| e^{-i \delta_2} \dpg + |{\cal A}_{2b}^{\pi\pi}| e^{-i \delta_3} \dg ,
\en
explicitly with the strong phases $\delta$'s and the weak phase $\gamma$, such as ${\cal E} = E e^{i\delta_E}$.
Here ${\cal T} (= T e^{i\delta_T})$ is taken to be real ({\it i.e.}, $\delta_T = 0$) and the other strong phases $\delta$'s are the relative ones to $\delta_T$.
We note that the weak phase $\gamma$ appears only in the terms having the CKM factor $\lambda_b$ or $\lambda^{(2)}_d$ of order $\lambda^5$.
Thus, one expects that the direct CP asymmetry for $D^0 \to \pi^+ \pi^-$ is CKM-suppressed by the factor of $\lambda^4 \sim 10^{-3}$.

\begin{table}
\caption{Topological amplitudes for $D \to PP$ ($P = \pi, K, \eta^{(\prime)}$) modes (singly Cabbibbo-suppressed decays). The strong phases $\delta$'s and the weak phase $\gamma$ are explicitly denoted. The CKM factor $ \lambda_d \equiv \lambda_d^{(1)}+\lambda_d^{(2)}$, where $\lambda_d^{(1)} = -\lambda +\frac{\lambda^3}{2} +{\lambda^5 \over 8} (1 +4 A^2)$ and
$\lambda_d^{(2)} = -\lambda^5 A^2 (\rho +i \eta) \equiv -|\lambda_2^{(2)}| e^{+i\gamma} = -|\lambda_b| e^{+i\gamma}$. For simplicity, singlet QCD penguin, flavor-singlet weak annihilation and electroweak-penguin amplitudes have been neglected.}
\label{t1}
\smallskip
\begin{tabular}{c c c} \hline \hline
 &  Mode ~~~~  &  Representation \\
\hline
$D \to$ & $\pi^+ \pi^-$ & $(\lambda^{(1)}_d - |\lambda^{(2)}_d| \dpg) (T + E \de + P_d \dpp + PA_d \dpa + PE_d \dpe) $ \\
  & & \qquad $+ \lambda_s (P_s \dpp + PA_s \dpa + PE_s \dpe) $ \\
  & & \qquad $+ |\lambda_b| (P_b \dpp + PA_b \dpa + PE_b \dpe)\dg$ \\

  & $\pi^0 \pi^0$ & ${1\over\sqrt{2}} [ (\lambda^{(1)}_d - |\lambda^{(2)}_d| \dpg) (-C \dc + E \de + P_d \dpp + PA_d \dpa+ PE_d \dpe) $ \\
  & & \qquad $+ \lambda_s (P_s \dpp+ PA_s \dpa+ PE_s \dpe)$ \\
  & & \qquad $+ |\lambda_b| (P_b \dpp+ PA_b \dpa+ PE_b\dpa)\dg ]$ \\

  & $\pi^+ \pi^0$ & ${1\over\sqrt{2}} (\lambda^{(1)}_d - |\lambda^{(2)}_d| \dpg) (T + C \dc)$  \\ \hline

  & $K^+ K^{-}$ & $(\lambda^{(1)}_d - |\lambda^{(2)}_d| \dpg) (P_d \dpp+ PA_d \dpa+ PE_d \dpe) $  \\
  & & \qquad $+ \lambda_s (T + E \de + P_s\dpp + PA_s \dpa+ PE_s \dpe)$ \\
  & & \qquad $+ |\lambda_b| (P_b \dpp+ PA_b \dpa+ PE_b \dpe) \dg$ \\

  & $K^0 \bar{K}^{0}$ &  $(\lambda^{(1)}_d - |\lambda^{(2)}_d| \dpg) (E \de+ 2 PA_d \dpa)$  \\
  & & \qquad $+ \lambda_s (E \de+ 2 PA_s \dpa)$   \\
  & & \qquad $+ |\lambda_b| (2 PA_b \dpa)\dg$ \\

  & $K^+ \bar{K}^{0}$ & $(\lambda^{(1)}_d - |\lambda^{(2)}_d| \dpg) (A  \da+P_d \dpp+ PE_d \dpe)$ \\
  & & \qquad $+ \lambda_s (T + P_s \dpp+ PE_s \dpe)$ \\
  & & \qquad $+ |\lambda_b| (P_b \dpp+ PE_b \dpe) \dg $  \\ \hline

  & $\pi^0 \eta$ & $(\lambda^{(1)}_d - |\lambda^{(2)}_d| \dpg) (-E \de+P_d\dpp+ PE_d\dpe) \cos\phi$  \\
  & & \qquad $- \lambda_s [ { 1\over\sqrt{2}} C \sin\phi \dc + (P_s \dpp+ PE_s\dpe)\cos\phi) ]$ \\
  & & \qquad $+ |\lambda_b| (P_b \dpp + PE_b \dpe) \cos\phi \dg$ \\

  & $\pi^0 \eta' $ & $(\lambda^{(1)}_d - |\lambda^{(2)}_d| \dpg) (-E \de+P_d \dpp+ PE_d \dpe) \sin\phi$  \\

  & $\pi^+ \eta $ & $(\lambda^{(1)}_d - |\lambda^{(2)}_d| \dpg) [ \frac{1}{\sqrt{2}}(T+C \dc+2A \da)+\sqrt{2}(P_d \dpp+PE_d \dpe)]\cos\phi$ \\
  & & \qquad $+ \lambda_s [- C\sin\phi \dc+\sqrt{2}(P_s \dpp+PE_s \dpe)\cos\phi]$ \\
  & & \qquad $+ |\lambda_b| [\sqrt{2}(P_b \dpp+PE_b \dpe)]\cos\phi \dg$  \\

  & $\pi^+ \eta' $ & $(\lambda^{(1)}_d - |\lambda^{(2)}_d| \dpg) [ \frac{1}{\sqrt{2}}(T+C \dc+2A \da)+\sqrt{2}(P_d \dpp+PE_d \dpe) ] \sin\phi$  \\
  & & + $\lambda_s [ C\cos\phi \dc+\sqrt{2}(P_s \dpp+PE_s \dpe)\sin\phi ]$  \\
  & & \qquad $+ |\lambda_b| [\sqrt{2}(P_b \dpp +PE_b \dpe) ] \sin\phi \dg$ \\
\hline \hline
       \end{tabular}
\end{table}
\begin{table}
\newpage
\caption{({\it Continued from Table I}) Topological amplitudes for $D_s \to PP$ ($P = \pi, K, \eta^{(\prime)}$) modes (singly Cabbibbo-suppressed decays).}
\label{t2}
\smallskip
\begin{tabular}{c c c} \hline \hline
 &  Mode ~~~~  &  Representation \\
\hline
$D_s \to$ & $\pi^+ K^{0}$ & $(\lambda^{(1)}_d - |\lambda^{(2)}_d| \dpg) (T +P_d \dpp+ PE_d \dpe)$  \\
  & & \qquad $+ \lambda_s (A \da+ P_s \dpp+ PE_s \dpe)$ \\
  & & \qquad $+ |\lambda_b| (P_b \dpp + PE_b \dpe) \dg $ \\

  & $K^{+}\pi^0$ & $\frac{1}{\sqrt{2}} (\lambda^{(1)}_d - |\lambda^{(2)}_d| \dpg) (-C \dc+P_d \dpp+ PE_d \dpe)$ \\
  & & \qquad $+ \lambda_s (A \da+ P_s \dpp+ PE_s \dpe)$ \\
  & & \qquad $+ |\lambda_b| (P_b \dpp+ PE_b \dpe) \dg$  \\ \hline

  & $K^{+}\eta$ & $(\lambda^{(1)}_d - |\lambda^{(2)}_d| \dpg) [\frac{1}{\sqrt{2}}(C \dc+P_d \dpp+ PE_d \dpe)\cos\phi -(P_d \dpp+ PE_d \dpe)\sin\phi]$  \\
  & & $+ \lambda_s [\frac{1}{\sqrt{2}}(A \da+P_d \dpp+ PE_d \dpe)\cos\phi$  \\
  & & \qquad \qquad \qquad $- (T+C \dc+A \da+P_s \dpp+ PE_s\dpe)\sin\phi]$ \\
  & & \qquad $+ |\lambda_b| (P_b \dpp +PE_b \dpe)(\frac{1}{\sqrt{2}}\cos\phi -\sin\phi ) \dg$ \\

  & $K^{+}\eta'$ & $(\lambda^{(1)}_d - |\lambda^{(2)}_d| \dpg) [\frac{1}{\sqrt{2}}(C \dc+P_d \dpp+ PE_d \dpe)\sin\phi +(P_d \dpp+ PE_d \dpe)\cos\phi]$ \\
  & & $\lambda_s [\frac{1}{\sqrt{2}}(A \da+P_d \dpp+ PE_d \dpe) \sin\phi$ \\
  & & \qquad \qquad \qquad $+(T+C \dc+A \da +P_s \dpp+ PE_s \dpe) \cos\phi]$ \\
  & & \qquad $|\lambda_b| (P_b \dpp +PE_b \dpe)(\frac{1}{\sqrt{2}}\sin\phi +\cos\phi) \dg$ \\
\hline \hline
\end{tabular}
\end{table}

Within the SM, for CP asymmetries in charm decays one has to go to the CKM matrix through order $\lambda^6$ in the Wolfenstein parametrization~\cite{Bigi:2011em} and understand the differences between other parameterizations~\cite{Ahn:2011fg}.
The SM generates CP asymmetries of order $\lambda^5$ for SCS charm decays and does not generate any CP ones for DCS decays up to order of $\lambda^6$.

In Table I and II, the decay amplitudes of $D \to PP$ ($P = \pi, K, \eta^{(\prime)}$) modes are expressed in terms of the topological amplitudes explicitly with the strong phases $\delta$'s and the weak phase $\gamma$.
For $D \to \pi \eta^{(\prime)}$ and $D_s \to K \eta^{(\prime)}$ modes, the $\eta-\eta'$ mixing is considered:
in the flavor basis,
\be
 \left(\matrix{ \eta \cr \eta'\cr}\right)=\left(\matrix{ \cos\phi & -\sin\phi \cr
 \sin\phi & \cos\phi\cr}\right)\left(\matrix{\eta_q \cr \eta_s
 \cr}\right),
\en
where $\eta_q={1\over\sqrt{2}}(u\bar u+d\bar d)$, $\eta_s=s\bar s$, and the mixing angle $\phi=40.4^\circ$~\cite{KLOE}.

In order to understand dynamics behind CP violation, one first needs to obtain decay amplitudes of the corresponding decays. The only model-independent way to analyze these amplitudes is to consider contributions from all possible quark diagram processes. We would like to point out that unlike all the previous works, we consider contributions from all possible strong penguin diagrams including $b$-quark processes. However, limited experimental information and the large number of parameters to be determined has made it a rather difficult task.

\section{The $\chi^2$ analysis and SU(3) breaking effects}

The topological amplitudes can be extracted from the available experimental information, such as branching fractions and direct CP asymmetries of SCS charm decays. Previous studies show that it is difficult to fit the experimental data in the $\chi^2$ analysis without using additional theoretical information such as QCDF and pQCD. It is because the number of parameters in theory are large and the experimental information particularly on $a_{CP}$ measurements in charm sector is rather poor. Also, the SU(3) breaking effects cannot be ignored, as it is impossible to fit the data within flavor SU(3) symmetry.

\begin{table}
\caption{Experimental data for $D \to PP$ ($P = \pi, K, \eta^{(\prime)}$) modes (singly Cabbibbo-suppressed decays). Branching fractions and direct CP asymmetries are shown in units of $10^{-3}$~\cite{Amhis:2012bh}. }
\label{t3}
\smallskip
\begin{tabular}{l c c} \hline \hline
 ~~~ Mode ~~~~  & ~~~~ Expt Br ($\times 10^{-3}$) ~~~~   & ~~~~ Expt $a_{CP}$ ($\times 10^{-3}$)   \\ \hline
 $D^0 \to \pi^+ \pi^-$       &  $1.400 \pm 0.026$  &  $2.2\pm 2.1$   \\
 $D^+ \to {\pi}^0{\pi}^+$    &  $1.19 \pm 0.05$   &  $29.0\pm 29.0$  \\
 $D^0 \to {\pi }^0{\pi }^0$  &  $0.82 \pm 0.05$  &  $0.0\pm 50.0$   \\
 $D^0 \to K^+K^-$            &  $3.96 \pm 0.08$  &  $-2.1\pm 1.7$   \\
 $D^+ \to \bar{K}^0K^+$            &  $5.66 \pm 0.32$  &  $-1.1\pm 2.5$   \\
 $D^0 \to \bar{K}^0K^0$            &  $0.34\pm 0.08$   & $230\pm 190$  \\
 $D^0 \to {\pi}^0\eta$       &  $0.68\pm 0.07$   &  $-$  \\
 $D^0 \to {\pi}^0\eta'$      &  $0.89\pm 0.14$   &  $-$  \\
 $D^+ \to {\pi}^+\eta $      &  $3.53\pm 0.21$   &  $10.0\pm 15.0$  \\
 $D^+ \to {\pi}^+\eta '$     &  $4.67\pm 0.29$   &  $-5.0\pm 12.0$  \\
 $D_s^+\to K^0{\pi}^+$       &  $2.42\pm 0.16$   &  $12.0\pm 10.0$  \\
 $D_s^+\to K^+{\pi}^0$       &  $0.63\pm 0.21$   &  $-266\pm 238$  \\
 $D_s^+\to K^+\eta $         &  $1.75\pm 0.35$   &  $93.0\pm 152$  \\
 $D_s^+\to K^+\eta'$         &  $1.80\pm 0.60$   &  $60.0\pm 189.0$  \\
\hline \hline
\end{tabular}
\end{table}

\subsection{The $\chi^2$ analysis in SU(3) limit}

We perform the $\chi^2$ analysis with 25 observables (experimental branching fractions and direct CP asymmetries) as inputs, shown in Table~\ref{t3}.
Within SU(3) symmetry, the total number of parameters is 20, including the magnitudes and strong phases of topological amplitudes, as shown in Table~\ref{t4}.
Although the degree of freedom (d.o.f.) in this case is 5, it turns out that it is impossible to obtain a fit with an acceptable $\chi^2/d.o.f.$
In this fit, we find $650 \leq \chi^2/d.o.f. \leq 1000$, being unacceptably large. This fact has also been supported by previous works based on a similar analysis~\cite{Cheng:2012wr}.

\subsection{The $\chi^2$ analysis including SU(3) breaking effects}

In order to include SU(3) breaking effects, we divide decay processes in categories of $\pi \pi,~ KK, ~ \pi \eta^{(')}, ~K\pi$ and $ K \eta^{(')}$ for $D$ and $D_s$ decays. Then we introduce five additional parameters in their amplitudes, namely, $\Delta_{\pi \pi},~ \Delta_{KK}, ~ \Delta_{\pi \eta}, ~\Delta_{K\pi}$ and $ \Delta_{K \eta}$, where each $\Delta_{PP}$ characterizes each category of $PP$ modes. The SU(3) broken decay amplitude in such case may, for example for $D^0 \to \pi^+ \pi^-$, be given as
\be
{\cal A}(D^0 \to \pi^+ \pi^-) &=& \big[(\lambda^{(1)}_d - |\lambda^{(2)}_d| \dpg) (T + E \de + P_d \dpp + PA_d \dpa + PE_d \dpe)
 \non \\
       &+& \lambda_s (P_s \dpp + PA_s \dpa + PE_s \dpe)  \non \\
       &+& |\lambda_b| (P_b \dpp + PA_b \dpa + PE_b \dpe)\dg~\big](1+\Delta_{\pi\pi}).
\en
It may be noted that introduction of SU(3) breaking parameters lowers the $ \chi^2/d.o.f. $ to acceptable limits.
To perform the $\chi^2$ analysis, we use 25 observables (experimental branching fractions and direct CP asymmetries) as inputs, shown in Table~\ref{t3}~\footnote{The CP asymmetry of $D^0 \to K^0 K^0$ has not been used as an input in the fit.}. The total maximum number of parameters in this analysis is 24, including the magnitudes and strong phases of topological amplitudes and SU(3) breaking parameters, as shown in Table~\ref{t4}. Based on the choice of parameters we have two cases as follows.

\begin{table}
\centering
\caption {{\it Case I} : Topological amplitudes and SU(3) breaking parameters determined from the fit {\it without any constraints}. The magnitudes and strong phases of the amplitudes are given in units of $10^{-6}$ GeV and degrees, respectively.}
\label{t4}
\smallskip
\renewcommand{\arraystretch}{1.2}
\begin{tabular}{c c c | c c c | c c c} \hline \hline
 ~ No. ~ & Parameter & ~ Value ~ & ~ No. ~ & Parameter & ~ Value ~ &  ~ No. ~ & Parameter & ~ Value  \\ \hline
 1  &  $T$    &  3.69  &  9   &  $PA_s$  &  1.29    &  17    &  $\delta_{P}$   &  $-144.10$  \\
 2  &  $C$    &  2.91  &  10  &  $PA_b$  &  1.47    &  18    &  $\delta_{PA}$   &  92.22   \\
 3  &  $A$    &  1.50  &  11  &  $PE_d$  &  0.30    &  19    &  $\delta_{PE}$  &  24.50   \\
 4  &  $E$    &  1.65  &  12  &  $PE_s$  &  2.46    &  20    &  $\Delta_{\pi\pi}$   &  0.006   \\
 5  &  $P_d$  &  2.18  &  13  &  $PE_b$  &  0.60    &  21    &  $\Delta_{K \pi}$    &  0.07    \\
 6  &  $P_s$  &  1.00  &  14  &  $\delta_{C}$  &  145.15   &  22 & $\Delta_{\pi \eta}$  & 0.21  \\
 7  &  $P_b$  &  1.35  &  15  &  $\delta_{A}$  &  $-143.53$ & 23 & $\Delta_{KK}$     &  0.011     \\
 8  &  $PA_d$ &  0.32  &  16  &  $\delta_{E}$  &  $-110.73$ & 24 & $\Delta_{K \eta}$ & 0.17  \\
 \hline \hline
\end{tabular}
\end{table}

\subsubsection{Case I: fit without any constraints}

We perform a complete analysis without using any constraints to extract 24 parameters with the weak phase $\gamma = 63^\circ$. In this case the degree of freedom is one. The obtained numerical values of all the parameters with $\chi^2/d.o.f. = 8.0$ are listed in Table~\ref{t4}~\footnote{
The $\chi^2/d.o.f.$ value does not seem small. However, this feature has been known in the previous works~\cite{Li:2012cfa,Bhattacharya:2012ah,Cheng:2010ry} using the similar global fit to $D \to PP$ data.
The $\chi^2 / d.o.f.$ value in the present fit mainly comes from the differences between theory fits and the central values of the poorly measured data for direct CP asymmetries. For the CP asymmetries, several modes in this fit induce sizable $\chi^2$ values: {\it e.g.,} for  $D^0 \to \pi^+ \pi^-$, $\pi^0 \pi^+$, $K^+ K^-$, $D_s \to K^0 \pi^+$, $K^+ \pi^0$, $K^+ \eta$, the corresponding $\chi^2$ values are 0.8, 1.0, 0.87, 1.4, 1.2, 0.8, respectively. }. For instance, it is found that
\be
&& {\cal T} = 3.69, ~~~~~~~~~~~~~~~~~~ {\cal C} = 2.91 e^{i 145.15^\circ},  \non \\
&& {\cal E} = 1.65 e^{-i 110.73^\circ}, ~~~~~~~ {\cal P}_d = 2.18 e^{-i 144.10^\circ}.
\en
Interestingly, the fitted values for ${\cal T, C, E}$ are quite close to the values obtained by other works~\cite{Cheng:2012wr,Bhattacharya:2012ah} for analysis of {\it Cabibbo-favored} modes of charm decays, except for the sign of the phases~\footnote{In our analysis we have carefully examined the value of $\chi^2/d.o.f.$ to determine the magnitudes and phases of the amplitudes. If the values of the phases change from the above ones, the $\chi^2/d.o.f.$ increases.}.
It is noted that topological amplitudes for the internal $b$-quark penguin are of the same order as the $s$- and $d$-quark penguin amplitudes, though $P_s \lesssim P_b < P_d$~\footnote{It may be argued that contributions from penguin amplitudes are usually expected to be considerably smaller in magnitude as compared to the color-allowed tree contribution. However, they may be enhanced due to long-distance FSI so as not to be neglected.}.

For $D \to \pi \pi, ~KK$ and $D_s \to K \pi$ modes, the SU(3) breaking effects are found to be only a few percents or less. For $D \to \pi \eta^{(')}$ and $D_s \to K \eta^{(')}$ modes, the SU(3) breaking becomes as large as about 20\%.

Using the fitted parameter values, the branching fractions and direct CP asymmetries are predicted, as shown in Table~\ref{t5}. We find that the fit obtained for the branching fractions is surprisingly good (Compare Table~\ref{t5} with Table~\ref{t3}.).
Almost all the predicted branching fractions are in very good agreement with the experimental data.
It should be emphasized that the branching fractions of $D^0 \to \pi^+{\pi}^-$ and $D^0 \to K^+ K^-$, having been a long standing puzzle, are also in excellent agreement with the data.
The only ambiguities in fitting arise for the $a_{CP}$'s of the observed decay modes which seems not to fit well in the present scenario. Since the experimental values of CP asymmetries still have very large uncertainties, in order to make a reliable conclusion, one has to wait for more precise CP measurements from the future experiments such as LHCb and Belle II.
We would like to remark that to the lowest order $a_{CP} (D \to PP)$ will include the $b$-quark penguin contribution multiplied by $\sin\gamma$ (which is close to 1 for  $\gamma=63^{\circ}$). Thus, the direct CP asymmetries of $D \to PP$ are expected to get a significant effect from the sizable contribution of the $b$-quark penguin shown in Table~\ref{t4}.

From the fitted parameters, one can make pure predictions for direct CP asymmetries of $D^0 \to K^0 K^0$, $\pi^0 \eta$ and $\pi^0 \eta'$ modes which have not been used as the inputs.  They are found to be
\be
&& a_{CP} (D^0 \to \bar{K}^0 K^0) = 1.42 \times 10^{-3},  \non \\
&& a_{CP} (D^0 \to \pi^0 \eta) = 2.8 \times 10^{-4},  \non \\
&& a_{CP} (D^0 \to \pi^0 \eta') = -6.8 \times 10^{-4}.
\en
The direct CP asymmetry for $D_s^+ \to K^+ \eta$ is expected to be large, being of $\mathcal{O}(10^{-2})$.  For $D^0 \to \bar{K}^0 K^0$ and $D^+ \to \pi^+ \eta$ modes, $a_{CP}$'s are expected to be of $\mathcal{O}(10^{-3})$.
For $D^0 \to \pi^0 \pi^0$, $\pi^0 \eta'$, $D^+ \to \bar{K}^0 K^+$ and $D_s^+\to K^+{\pi}^0$ modes, $a_{CP}$'s are close to $\mathcal{O}(10^{-3})$.
For the other modes including $D^0 \to \pi^+ \pi^-$, $a_{CP}$'s are of $\mathcal{O}(10^{-4})$ or smaller.
Future experimental measurements for these observables will provide useful information on this scenario.

\begin{table}
\caption {{\it Case I} : Branching fractions and direct CP asymmetries (in units of $10^{-3}$) obtained from the fit {\it without any constraints}.}
\label{t5}
\smallskip
\begin{tabular}{l c c } \hline \hline
 ~~~ Mode ~~~ & ~~~ Br ($\times 10^{-3}$) ~~~  & ~~~ $a_{CP}$ ($\times 10^{-3}$)  \\ \hline
 $D^0 \to \pi^+{\pi}^-$   &  $1.40$  &  0.30   \\
 $D^+ \to {\pi}^0{\pi}^+$     &  1.19    &  0      \\
 $D^0 \to {\pi}^0{\pi}^0$     &  0.83    &  0.86   \\
 $D^0 \to K^+ K^-$            &  3.96    &  $-0.51$  \\
 $D^+ \to \bar{K}^0 K^+$            &  5.68    &  $-0.76$  \\
 $D^0 \to \bar{K}^0 K^0$            &  0.36    &  1.42   \\
 $D^0 \to {\pi}^0\eta$        &  0.68    &  0.28   \\
 $D^0 \to {\pi}^0\eta'$       &  0.89    &  $-0.68$  \\
 $D^+ \to {\pi}^+\eta$        &  3.50    &  1.50   \\
 $D^+ \to {\pi}^+ \eta'$      &  4.65    &  $-1.02$  \\
 $D_s^+\to K^0{\pi}^+$        &  2.41    &  0.30   \\
 $D_s^+\to K^+{\pi}^0$        &  0.66    &  $-0.93$  \\
 $D_s^+\to K^+ \eta $         &  1.56    &  $-42.0$    \\
 $D_s^+\to K^+ \eta'$         &  2.22    &  $0.47$   \\
\hline \hline
\end{tabular}
\end{table}

\begin{table}
\centering
\caption {{\it Case II} : Topological amplitudes and SU(3) breaking parameters determined from the fit {\it with} $C = 0.8 T$ and $P_s = (1.0\pm 0.2 ) P_d$. The magnitudes and strong phases of the amplitudes are given in units of $10^{-6}$ GeV and degrees, respectively.}
\label{t6}
\smallskip
\renewcommand{\arraystretch}{1.2}
\begin{tabular}{c c c | c c c | c c c} \hline \hline
 ~ No. ~ & Parameter & ~ Value ~ & ~ No. ~ & Parameter & ~ Value ~ &  ~ No. ~ & Parameter & ~ Value  \\ \hline
 1  &  $T$    &  3.13  &  9   &  $PA_s$  &  1.01    &  17  &  $\delta_{P}$  &  $-145.11$   \\
 2  &  $C$    &  2.50  &  10  &  $PA_b$  &  1.48    &  18  &  $\delta_{PA}$  &  81.46    \\
 3  &  $A$    &  1.26  &  11  &  $PE_d$  &  0.52    &  19  &  $\delta_{PE}$ &  21.65   \\
 4  &  $E$    &  1.96  &  12  &  $PE_s$  &  2.42    &  20  &  $\Delta_{\pi\pi}$  &  0.003   \\
 5  &  $P_d$  &  1.84  &  13  &  $PE_b$  &  0.50    &  21  &  $\Delta_{K \pi}$   &  0.35    \\
 6  &  $P_s$  &  1.84  &  14  &  $\delta_{C}$  &  137.54  &  22  &  $\Delta_{\pi \eta}$  &  0.35  \\
 7  &  $P_b$  &  1.50  &  15  &  $\delta_{A}$  &  $-159.66$  &  23  &  $\Delta_{KK}$  &  0.33  \\
 8  &  $PA_d$ &  0.92  &  16  &  $\delta_{E}$ &  $-113.50$  &  24  &  $\Delta_{K \eta}$  & 0.32  \\
 \hline \hline
\end{tabular}
\end{table}

\subsubsection{Case II: fit with $C = 0.8 T$ and $P_s = (1.0\pm 0.2 ) P_d$}

We put some constraints on our analysis encouraged by the fit obtained by~\cite{Cheng:2012wr} for {\it Cabibbo-favored} decay modes of charmed mesons. With the constraints $C = 0.8 T$ and $P_s = (1.0 \pm 0.2 ) P_d$, we obtain the fit for the remaining 22 parameters. The relation $C = 0.8 T$ holds also in the result of Case I.
In this case the degree of freedom is 3. The fitted values of the parameters with $\chi^2/d.o.f. = 9.6$ are shown in Table~\ref{t6}.  Though the parameter values are similar to those obtained in Case I, in this case $P_b \approx P_d = P_s$, and the SU(3) breaking effects are as large as $32 \sim 35\%$ for all the modes, except $\pi \pi$ modes where the SU(3) breaking turns out to be very small.

Table~\ref{t7} shows the predicted branching fractions and $a_{CP}$'s. Here also the fit for branching fractions is excellent.
As in Case I, the branching fractions of $D^0 \to \pi^+{\pi}^-$ and $D^0 \to K^+ K^-$ are also in excellent agreement with the experimental data.
In this case, compared with Case I, more pure predictions for direct CP asymmetries can be made.  From the fitted values of the parameters, we find for $D^0 \to \pi^+ \pi^-$, $K^+ K^-$, $\bar{K}^0 K^0$, $\pi^0 \eta$ and $\pi^0 \eta'$ modes which have not been used as the inputs,
\be
&& a_{CP} (D^0 \to \pi^+ \pi^-) = 3.3 \times 10^{-4},  \non \\
&& a_{CP} (D^0 \to K^+ K^-) = -7.3 \times 10^{-4},  \non \\
&& a_{CP} (D^0 \to \bar{K}^0 K^0) = -5.8 \times 10^{-4},  \non \\
&& a_{CP} (D^0 \to \pi^0 \eta) = -4.7 \times 10^{-4},  \non \\
&& a_{CP} (D^0 \to \pi^0 \eta') = -3.3 \times 10^{-4}.
\en
In this case, the direct CP asymmetries for several modes are expected to be smaller, compared with Case I.
The direct CP asymmetries for $D_s^+ \to K^+\pi^0$ and $K^+ \eta$ are of $\mathcal{O}(10^{-3})$.
For $D^0 \to{\pi }^0{\pi}^0$, $K^+K^-$, $D^+ \to {\pi}^+ \eta$, ${\pi}^+ \eta'$, $a_{CP}$ are close to $\mathcal{O}(10^{-3})$.
For the other modes including $D^0 \to \pi^+ \pi^-$, $a_{CP}$'s are expected to be of $\mathcal{O}(10^{-4})$ or smaller.
Future experiments such as upgraded LHCb and Belle II will help to determine which scenario is more reliable.

\begin{table}
\caption {{\it Case II} : Branching fractions and direct CP asymmetries (in units of $10^{-3}$) obtained from the fit {\it with} $C = 0.8 T$ and $P_s = (1.0 \pm 0.2 ) P_d$.}
\label{t7}
\smallskip
\begin{tabular}{l c c } \hline \hline
 ~~~ Mode ~~~ & ~~~ Br ($\times 10^{-3}$) ~~~  & ~~~ $a_{CP}$ ($\times 10^{-3}$)  \\ \hline
 $D^0 \to \pi^+{\pi}^-$  &  1.40  &  0.33   \\
 $D^+ \to{\pi }^0{\pi}^+$    &  1.18  &  0      \\
 $D^0 \to{\pi }^0{\pi}^0$    &  0.89  &  0.71   \\
 $D^0 \to K^+K^-$            &  3.98  &  $-0.73$  \\
 $D^+ \to \bar{K}^0K^+$            &  5.56  &  $-0.17$  \\
 $D^0 \to \bar{K}^0K^0$            &  0.39  &  $-0.58$   \\
 $ D^0 \to {\pi}^0 \eta$     &  0.68  &  $-0.47$  \\
 $D^0 \to {\pi}^0 \eta'$     &  0.91  &  $-0.33$   \\
 $D^+ \to {\pi}^+ \eta $     &  3.39  &  0.89   \\
 $D^+ \to {\pi}^+ \eta'$     &  4.57  &  $-0.67$  \\
 $D_s^+ \to K^0 \pi^+$       &  2.41  &  $-0.15$  \\
 $D_s^+ \to  K^+\pi^0$       &  0.54  &  $-2.43$  \\
 $D_s^+ \to K^+\eta $        &  1.48  &  1.4   \\
 $D_s^+ \to K^+ \eta'$       &  2.32  &  $-0.23$  \\
\hline \hline
\end{tabular}
\end{table}

\section{Conclusions}

We have performed a model-independent analysis of CP violation in the singly Cabibbo-suppressed $D \to PP$ decays. In light of the most recent \CP asymmetry measurements $a_{CP}^{\rm dir}$ for the $D^0 \to K^+ K^-$ and $D^0 \to \pi^+ \pi^-$ modes, we have  analyzed direct \CP violation in $D \to PP$ decays within the SM.
It is believed that direct CP violation in such decays may arise from the interference between various topological amplitudes and the asymmetries are expected to be of $\mathcal{O}(10^{-3})$ or smaller.
We have taken advantage of the quark diagram approach to extract the topological amplitudes and relative strong phases from the measured experimental data.
We have also taken care of SU(3) breaking effects which are considered to be large in these decays. It may be noted that for charmed meson decays it is difficult to induce CP violation at tree level. Thus, one has to include contributions from penguin amplitudes at certain level.
In order to explain the experimental branching fractions, we have included contributions from all possible strong penguin amplitudes involving penguin exchange and penguin annihilation diagrams.
We have already stated that due to the large uncertainties in measured $a_{CP}$'s, CP violation in charmed meson decays has not yet established experimentally. This indicates that a more precise approach is needed to analyze CP violation in these decays. Therefore, we have included all the possible internal $b$-quark contributions to penguin diagrams (up to order of $\lambda^6$)  which have been ignored so far in the previous works.

In the quark diagram approach, we have determined topological amplitudes for $D \to PP$ decays through the $\chi^2$ analysis by using the available experimental branching fractions and direct CP asymmetries.
Our $\chi^2$ analysis has been performed in the most general way, {\it i.e., without using any model-dependent information} on the parameters (Case I).
We have divided our analysis into two cases: without any constraints (Case I) and with certain constraints (Case II), having the best $\chi^2 / d.o.f.$ = $8.0$ and $9.6$, respectively. Consequently, we have predicted the unmeasured direct CP asymmetries. Our findings are summarized as follows.
\begin{enumerate}
\item We observe an excellent fit for branching fractions, including ${\cal B} (D^0 \to \pi^+ \pi^-)$ and ${\cal B}(D^0 \to K^+ K^-)$, in both the cases as shown in tables \ref{t5} and \ref{t7}.
\item It may be argued that contributions from penguin amplitudes are considerably smaller in magnitude as compared to the color-allowed tree contribution. However, they may be enhanced due to long-distance FSI resonances so as not to be neglected.
\item We find that penguin contributions from the internal $b$-quark can be sizable and especially for CP asymmetries in charm decays, they become important and cannot be neglected.
\item Inclusion of $b$-quark contributions result in CP asymmetries that lies in range within the SM: \textit{i.e.,} $10^{-4} \leq a_{CP} \leq 10^{-3}$ for both the cases.
\item In \textit{Case I}: (i) The observed $\mathcal{T}, \mathcal{C}$ and $\mathcal{E}$ acquire values closer to those found in a similar analysis based on Cabibbo-favored charm decays. The SU(3) breaking is up to $20\% $ for $\pi \eta^{(')}$ and $K \eta^{(')}$ modes. \\
    (ii) We predict $a_{CP}$'s for $D^0 \to \bar{K}^0 K^0/\pi\eta/\pi\eta'$ modes to be of $\mathcal{O}(10^{-3}) \sim \mathcal{O}(10^{-4})$.
\item In \textit{Case II}: (i) We use constraints $C = 0.8 T$ and $P_s = (1.0 \pm 0.2 ) P_d$ to increase our predictability and to test our fit. We predict $a_{CP} (\pi^{+}\pi^{-}) = 0.33 \times 10^{-3}$ and $a_{CP} (K^{+}K^{-}) =-0.73 \times 10^{-3}$ with $\Delta a_{CP}=-1.00\times 10^{-3}$, which are consistent with the recent experimental result. \\
    (ii) We also predict $a_{CP}$'s for $\bar{K}^0 K^0/\pi\eta/\pi\eta'$ modes to be of the same order of magnitude and sign.  \\
    (iii) SU(3) symmetry breaking as large as $35\%$ is observed in Case II for the modes other than $\pi \pi$.
\end{enumerate}
We wish to remark here that our results are more or less comparable/consistent with other theoretical works based on similar approaches. In order to get a clearer picture, more precise experimental measurements are required to determine the exact magnitude of CP asymmetries in charmed meson decays. We hope that such an analysis would be helpful in diagnosing possible evidence of new physics in the charm sector.
\\

\noindent
{\it * Note added:} When finishing the paper, we have found that two new works~\cite{Muller:2015lua,Biswas:2015aaa} have just come out. Although those two works also use the similar approach to ours, there are clear differences between our work and theirs.
As mentioned in Introduction and Conclusions, we have performed the $\chi^2$ analysis in the most general way, {\it i.e., without using any model-dependent information} on the parameters.
In order to get more precise and reliable estimates of the amplitudes, we have included contributions from all possible strong penguin amplitudes, including the internal $b$-quark penguins.
Then, we have made predictions for direct CP asymmetries of SCS $D$ decay modes, including yet unmeasured ones.


\acknowledgments \vspace{-2ex}
We thank Yeo Woong Yoon for useful comments.
The work is supported by the National Research Foundation of Korea (NRF) grant
funded by Korea government of the Ministry of Education, Science and
Technology (MEST) (Grant No. 2011-0017430) and (Grant No. 2011-0020333).
\\


\end{document}